\documentclass[3p,times]{elsarticle}

\usepackage{amsfonts}
\usepackage{mathrsfs}
\usepackage{amsmath}
\usepackage{amssymb}

 \usepackage{lineno}
\newtheorem{theorem}{\indent Theorem}[section]
\newtheorem{corollary}{\indent Corollary}[section]
\newtheorem{proposition}{\indent Proposition}[section]
\newtheorem{definition}{\indent Definition}[section]
\newtheorem{lemma}{\indent Lemma}[section]

\begin{document}

\begin{frontmatter}



\title{Dynamic of the three dimensional viscous primitive equations of large-scale atmosphere}


\author{Bo You\corref{cor1}}
\ead{youb2013@xjtu.edu.cn}
\cortext[cor1]{Corresponding author}
 \address{School of Mathematics and Statistics, Xi'an Jiaotong University\\
  Xi'an, 710049, P. R. China}

\begin{abstract}
The main objective of this paper is to study the existence of a finite dimension global attractor for the three dimensional viscous primitive equations of large-scale atmosphere. Thanks to the shortage of the uniqueness of weak solutions, we prove the existence of a global
attractor with finite fractal dimension for the three dimensional viscous primitive equations of large-scale atmosphere by using the method of $\ell$-trajectories.
\end{abstract}


\end{frontmatter}

\section{Introduction}
\def\theequation{1.\arabic{equation}}\makeatother
\setcounter{equation}{0}
 In this paper, we consider the long-time behavior of solutions for the following three dimensional viscous primitive equations of large-scale atmosphere(see \cite{ljl, ljl1}):
 \begin{equation} \label{1.1}
 \begin{cases}
&\frac{\partial v}{\partial t}+(v\cdot \nabla)v+w\frac{\partial v}{\partial z}+f_0\vec{k} \times v+\nabla p+ L_1v=0,\\
&\frac{\partial p}{\partial z}=T,\\
&\nabla\cdot v+\frac{\partial w}{\partial z}=0,\\
&\frac{\partial T}{\partial t}+v\cdot\nabla T+w\frac{\partial T}{\partial z}+L_2T=Q
\end{cases}
\end{equation}
in the domain
\begin{align*}
\Omega=M\times (-h,0)\subset \mathbb{R}^3,
\end{align*}
where $M\subset\mathbb{R}^2$ is a bounded domain with smooth
boundary. Here $v=(v_1,v_2)$, $(v_1,v_2,w)$ is the velocity field,
$T$ is the temperature, $p$ is the pressure, $f_0 =
2\Omega_1\sin\nu_0$ is the Coriolis parameter, $\vec{k}=(0,0,1)$ is
vertical unit vector and $Q$ is a heat
source. The operators $L_1$ and $L_2$ are given by
\begin{align*}
&L_1=-\frac{1}{Re_1}\Delta-\frac{1}{Re_2}\frac{\partial^2}{\partial z^2},\\
&L_2=-\frac{1}{Rt_1}\Delta-\frac{1}{Rt_2}\frac{\partial^2}{\partial
z^2},
\end{align*}
where $Re_1,$ $Re_2$ are positive constants representing the
horizontal and vertical Reynolds numbers, respectively, and $Rt_1,$
$Rt_2$ are positive constants which stand for the horizontal and
vertical heat diffusivity, respectively. For the sake of simplicity, let
$\nabla=(\partial_x,\partial_y)$ be the horizontal gradient operator
and let $\Delta=\partial^2_x+\partial^2_y$ be the horizontal
Laplacian. We denote the different parts of the boundary of $\bar{\Omega}$ by
\begin{align*}
&\Gamma_u=\{(x,y,z)\in \bar{\Omega}:z=0\},\\
&\Gamma_b=\{(x,y,z)\in \bar{\Omega}:z=-h\},\\
&\Gamma_l=\{(x,y,z)\in \bar{\Omega}:(x,y)\in \partial M,-h\leq z
\leq 0\}.
\end{align*}
Equations \eqref{1.1} is equipped with the following boundary
conditions, with non-slip and non-flux on the side walls and bottom
(see \cite{ccs1})
\begin{equation}\label{1.2}
\begin{cases}
&\frac{\partial v}{\partial z}|_{\Gamma_u}=0,w|_{\Gamma_u}=0,(\frac{1}{Rt_2}\frac{\partial T}{\partial z}+\alpha T)|_{\Gamma_u}=0,\\
&\frac{\partial v}{\partial z}|_{\Gamma_b}=0,w|_{\Gamma_b}=0, \frac{\partial T}{\partial z}|_{\Gamma_b}=0,\\
&v\cdot\vec{n}|_{\Gamma_l}=0,\frac{\partial
v}{\partial\vec{n}}\times \vec{n}|_{\Gamma_l}=0,\frac{\partial
T}{\partial \vec{n}}|_{\Gamma_l}=0.
\end{cases}
\end{equation}
In addition, we supply equations \eqref{1.1}-\eqref{1.2} with the following initial datum
\begin{equation}\label{1.3}
\begin{cases}
&v(x,y,z,0)=v_0(x,y,z),\\
&T(x,y,z,0)=T_0(x,y,z).
\end{cases}
\end{equation}
 In the past several decades, the primitive
equations of the atmosphere, the ocean and the coupled
atmosphere-ocean have been extensively studied from the mathematical
point of view (see \cite{ccs1, ebd, gbl2,gbl, gbl1, hc1,hc, ljl, ljl1, mtt,
pm} etc). By introducing $p$-coordinate system and using some technical treatments, Lions, Temam and Wang in \cite{ljl} obtained a new formulation for the
primitive equations of large-scale dry atmosphere which is a little similar with Navier-Stokes
equations of incompressible fluid, and they proved the existence of
weak solutions for the primitive equations of the atmosphere. In
\cite{ljl1}, Lions, Temam and Wang introduced the primitive equations of large-scale ocean and proved the existence of weak solutions and the well-posedness of local in time strong solutions for the primitive equations of large-scale ocean, and estimated the dimension of the universal attractor. Based on the works of
Lions, Temam and Wang in \cite{ljl, ljl1}, many authors continued to consider the well-posedness of solutions for the primitive equations of large-scale atmosphere (see \cite{ccs3,ccs4,ccs2, ebd, gf,gbl2, gbl3, hc1,hc, mtt, tr1,zmc1, zmc}). However, the uniqueness of weak solutions and the global existence of strong solutions for the three dimensional primitive equations of large-scale ocean and atmosphere dynamics with any initial datum remain unresolved. Until 2007, Cao and Titi \cite{ccs1} decomposed the three dimensional primitive equations of large-scale ocean and atmosphere dynamics into two systems by using the idea of the decomposition of semigroup, one is similar with the two dimensional incompressible Navier-Stokes equations, the other is the reaction-convection-diffusion equations. As we known, the solutions of each system were fairly regular. Cao and Titi performed some a priori estimates about the solutions of each system by which they obtained some a priori estimates of strong solutions for the three dimensional primitive equations of large-scale ocean and atmosphere dynamics, which implies the well-posedness of strong solutions for the three dimensional primitive equations of large-scale ocean and atmosphere dynamics, they resolved the open question posed in \cite{ljl, ljl1}. Meanwhile, the long-time behavior of solutions for the three dimensional primitive equations of large-scale ocean and atmosphere dynamics has been considered extensively (see \cite{ci, elc, gbl2,gbl, gbl1, gbl3, hng,hch, jn, jn2, jn1, lk, yb2}). In particular, in \cite{gbl}, Guo and Huang obtained a weakly compact global attractor $\mathcal{A}$ for the primitive equations of large-scale atmosphere which captures all the trajectories. The existence of a global attractor in $V$ for the primitive equations of large-scale atmosphere and ocean dynamics was proved by Ning Ju in \cite{jn} by using the Aubin-Lions compactness theorem under the assumption $Q\in L^2(\Omega).$ In \cite{jn2, jn1}, the authors have proved the finite dimensional global attractor for the 3D viscous primitive equations by using the squeezing property.

 To the best of our knowledge, the method of $\ell$-trajectories is based on an observation
that the limit behavior of solutions to a dynamical system in an original
phase space can be equivalently captured by the limit behavior of
$\ell$-trajectories which are continuous parts of solution trajectories that are
para-metrized by time from an interval of the length $\ell$ with $\ell>0$ and it can weaken the requirements on
the regularity of the solution. In this paper, thanks to the shortage of the regularity of weak solutions for the three dimensional viscous primitive equations of large-scale atmosphere, we can not obtain the uniqueness of weak solutions such that we are not able to define the semigroup on $H.$ Therefore, the existence of a global attractor in $H$ can not be obtained by the classical theory of dynamical systems. Fortunately, we know that any weak solutions of equations \eqref{1.1}-\eqref{1.3} will be unique for any $t>0.$  To overcome this difficulty, inspired by the idea of the method of $\ell$-trajectories for any small $\ell>0$ proposed in \cite{mj}, in this paper, we first define a semigroup $\{L_t\}_{t\geq 0}$ on some subset $X_\ell$ of $L^2(0,\ell;H)$ generated by problem \eqref{1.1}-\eqref{1.3}, and then, we prove the existence of a global attractor $\mathcal{A}_\ell$ in $X_\ell$ for the semigroup $\{L_t\}_{t\geq 0}$ by the method of $\ell$-trajectories and estimate the fractal dimension of the global attractor by using the smooth property of the difference of two solutions. Finally, by defining a Lipschitz continuous operator on the global attractor $\mathcal{A}_\ell,$ we obtain the existence of a finite dimensional global attractor $\mathcal{A}$ in the original phase space $H$ for problem \eqref{1.1}-\eqref{1.3}.

Throughout this paper, let $C$ be a generic constant that is independent of the initial datum of $(v,T).$
\section{Mathematical setting of equations \eqref{1.1}-\eqref{1.3}}
\def\theequation{2.\arabic{equation}}\makeatother
\setcounter{equation}{0}
\subsection{Reformulation of equations \eqref{1.1}-\eqref{1.3}}
Integrating the third equation of \eqref{1.1} and combining the boundary condition \eqref{1.2}, we obtain
\begin{align*}
w(x,y,z,t)=-\int_{-h}^z \nabla\cdot v(x,y,r,t)dr
\end{align*}
and
\begin{align}
\label{2.1}\int_{-h}^0\nabla\cdot v(x,y,r,t)dr=0.
\end{align}
 Define an unknown function on $z=0,$ say
 \begin{align*}
 p_s: M\rightarrow \mathbb {R},
 \end{align*}
 which is the pressure of the atmosphere on $z=0.$ Then
\begin{align}
\label{2.2}p(x,y,z,t)=p_s(x,y,t)+\int_0^zT(x,y,r,t)dr.
\end{align}
 Therefore, equations \eqref{1.1} can be reformulated as follows
\begin{equation}\label{2.3}
\begin{cases}
&\frac{\partial v}{\partial t}+(v\cdot\nabla)v-(\int_{-h}^z \nabla\cdot v(x,y,s,t)ds)\frac{\partial v}{\partial z}+\nabla p_s(x,y,t)+f_0\vec{k} \times v+\int_0^z \nabla T(x,y,s,t)ds+L_1v=0,\\
&\frac{\partial T}{\partial t}+v\cdot\nabla T-(\int_{-h}^z\nabla\cdot v(x,y,s,t)ds)\frac{\partial
T}{\partial z}+L_2 T=Q
\end{cases}
\end{equation}
subject to the following boundary conditions
\begin{equation}\label{2.4}
\begin{cases}
&\frac{\partial v}{\partial
z}|_{\Gamma_b}=0,\frac{\partial v}{\partial
z}|_{\Gamma_u}=0,v\cdot\vec{n}|_{\Gamma_l}=0,\frac{\partial
v}{\partial\vec{n}}\times \vec{n}|_{\Gamma_l}=0,\\
&\frac{\partial T}{\partial
z}|_{\Gamma_b}=0,(\frac{1}{Rt_2}\frac{\partial T}{\partial z}+\alpha
T)|_{\Gamma_u}=0,\frac{\partial T}{\partial \vec{n}}|_{\Gamma_l}=0
\end{cases}
\end{equation}
and the initial datum
\begin{equation}\label{2.5}
\begin{cases}
 &v(x,y,z,0)=v_0(x,y,z),\\
&T(x,y,z,0)=T_0(x,y,z).
\end{cases}
\end{equation}
\subsection{Some function spaces}
In this subsection, we first introduce the notations for some standard
function spaces on $\bar{\Omega}$ as follows
\begin{align*}
\mathcal {V}_1=&\left\{v\in(C^\infty(\bar{\Omega}))^2:\frac{\partial
v}{\partial z}|_{\Gamma_u}=0,\frac{\partial v}{\partial
z}|_{\Gamma_b}=0,v\cdot\vec{n}|_{\Gamma_l}=0,\frac{\partial
v}{\partial\vec{n}}\times \vec{n}|_{\Gamma_l}=0,\int_{-h}^0\nabla\cdot v(x,y,z)\,dz=0\right\},\\
 \mathcal{V}_2=&\left\{ T\in C^\infty(\bar{\Omega}):\frac{\partial T}{\partial
z}|_{\Gamma_b}=0,(\frac{1}{Rt_2}\frac{\partial T}{\partial z}+\alpha
T)|_{\Gamma_u}=0,\frac{\partial T}{\partial
\vec{n}}|_{\Gamma_l}=0\right\}.
\end{align*}
Denote the norm in $L^p(\Omega)(1\leq p<\infty)$ by the notation $\|\cdot\|_p$ given by
\begin{align*}
 \|h\|_p=\left(\int_\Omega |h|^p\,dxdydz\right)^{\frac{1}{p}}
\end{align*}
for any $h\in L^p(\Omega),$ and let $H_1$ be the closure of $\mathcal{V}_1$ with respect to the $(L^2(\Omega))^2$-norm. Similarly, we let $V_1,$
$V_2$ be the closure of $\mathcal{V}_1,$ $\mathcal{V}_2$
with respect to the following norms
\begin{align*}
\|v\|=&\left(\frac{1}{Re_1}\int_\Omega|\nabla v|^2\,dxdydz+\frac{1}{Re_2}\int_\Omega|\frac{\partial v}{\partial z}|^2\,dxdydz\right)^{\frac{1}{2}},\\
\|T\|=&\left(\frac{1}{Rt_1}\int_\Omega|\nabla T|^2\,dxdydz+\frac{1}{Rt_2}\int_\Omega|\frac{\partial T}{\partial z}|^2\,dxdydz+\alpha\int_{\Gamma_u} |T|^2\,dxdy\right)^{\frac{1}{2}}
\end{align*}
for any $v\in \mathcal{V}_1$ and $T\in
\mathcal{V}_2,$ respectively, let $V=V_1\times V_2$ and $H=H_1\times L^2(\Omega)$ with the norm defined by $\|(v,T)\|_2=\left(\|v\|_2^2+\|T\|_2^2\right)^{\frac{1}{2}}$ for any $(v,T)\in H,$ and denote by $V^{'}$ the dual space of $V.$

 Next, we recall some results used to prove the existence of a finite dimensional global attractor for problem \eqref{2.3}-\eqref{2.5}.
 \begin{lemma}\label{2.2.1}(\cite{ccs1})
There exists a positive constant $K_1$ such that
\begin{align*}
\frac{1}{K_1}\|T\|^2\leq\|T\|_{H^1(\Omega)}^2\leq K_1\|T\|^2
\end{align*}
for any $T\in V_2.$ Moreover, we have
\begin{align*}
\|T\|^2_2\leq K_2\|T\|^2
\end{align*}
for any $T\in V_2,$ where
\begin{align*}
K_2=\max\{\frac{2h}{\alpha},2Rt_2h^2\}.
\end{align*}
\end{lemma}

\begin{lemma}(\cite{cvv,jn, mj, mj1,sj2})\label{2..2.2}
Assume that $p_1\in(1,\infty],$ $p_2\in [1,\infty).$ Let $X$ be a Banach space and let $X_0,$ $X_1$ be separable and reflexive Banach spaces such that $X_0\subset\subset X\subset X_1.$ Then
\begin{align*}
Y=\{u\in L^{p_1}(0,\ell;X_0):u'\in L^{p_2}(0,\ell;X_1)\}\subset\subset L^{p_1}(0,\ell;X),
\end{align*}
where $\ell$ is a fixed positive constant.
\end{lemma}

 \begin{definition}(\cite{rjc,tr})
Let $\{S(t)\}_{t\geq0}$ be a semigroup on a Banach space $X.$ A set $\mathcal{A}\subset X$ is said to be a global attractor if the following conditions hold:
\begin{itemize}
\item [(i)] $\mathcal{A}$ is compact in $X.$
\item [(ii)] $\mathcal{A}$ is strictly invariant, i.e., $S(t)\mathcal{A}=\mathcal{A}$ for any $t\geq 0.$
\item [(iii)] For any bounded subset $B\subset X$ and for any neighborhood $\mathcal{O}=\mathcal{O}(\mathcal{A})$ of $\mathcal{A}$ in $X,$
there exists a time $\tau_0=\tau_0(B)$ such that $S(t)B\subset\mathcal{O}(\mathcal{A})$ for any $t\geq \tau_0.$
\end{itemize}
\end{definition}
\begin{lemma}(\cite{mj1})\label{2.2.3}
Let $X$ be a (subset of) Banach space and $(S(t),X)$ be a dynamical system. Assume that there exists a compact set $K\subset X$ which is uniformly absorbing and positively invariant with respect to $S(t).$ Let moreover $S(t)$ be continuous on $K.$ Then $(S(t),X)$ has a global attractor.
\end{lemma}
\begin{definition}(\cite{rjc, tr})
Let $H$ be a separable real Hilbert space. For any non-empty compact subset $K\subset H,$ the fractal dimension of $K$ is the number
\begin{align*}
d_f(K)=\limsup_{\epsilon\rightarrow 0^+}\frac{\log(N_\epsilon(K))}{\log(\frac{1}{\epsilon})},
\end{align*}
where $N_\epsilon(K)$ denotes the minimum number of open balls in $H$ with radii $\epsilon>0$ that are necessary to cover $K.$
\end{definition}
\begin{lemma}(\cite{mj1})\label{2.2.4}
Let $X,$ $Y$ be norm spaces such that $X\subset\subset Y$ and $\mathcal{A}\subset Y$ be bounded. Assume that there exists a mapping $L$ such that $L\mathcal{A}=\mathcal{A}$ and $L:Y\rightarrow X$ is Lipschitz continuous on $\mathcal{A}.$ Then $d_f(\mathcal{A})$ is finite.
\end{lemma}
\begin{lemma}(\cite{mj1})\label{2.2.5}
Let $X$ and $Y$ be two metric spaces and $f:X\rightarrow Y$ be $\alpha$-H\"{o}lder continuous on the subset $A\subset X.$ Then
\begin{align*}
d_F(f(A),Y)\leq\frac{1}{\alpha}d_F(A,X).
\end{align*}
In particular, the fractal dimension does not increase under a Lipschitz continuous mapping.
\end{lemma}
%
%
 \section{The existence of a global attractor}
 \def\theequation{3.\arabic{equation}}\makeatother
\setcounter{equation}{0}
We start with the following general existence
of weak solutions which can be obtained by the standard Faedo-Galerkin
methods (see \cite{ccs1, tr}). Here we only state
the result as follows.
\begin{theorem}\label{3.1}
Assume that $Q\in L^2(\Omega).$ Then for any $(v_0,T_0)\in H,$ there exists at least one solution $(v(t),T(t))\in \mathcal{C}(\mathbb{R}^+;H_w)\cap L_{loc}^2(\mathbb{R}^+;V)$ of problem \eqref{2.3}-\eqref{2.5}.
\end{theorem}

\begin{lemma}(\cite{ccs1, jn})\label{3.2}
Assume that $Q\in L^2(\Omega).$ Then for any $(v_0,T_0)\in V,$ there exists a unique strong solution $(v(t),T(t))\in \mathcal{C}(\mathbb{R}^+;V)$ of problem \eqref{2.3}-\eqref{2.5}, which depends continuously on the initial data with respect to the topology of $H$ and the topology of $V.$
\end{lemma}

\begin{corollary}\label{3.3}
Assume that $Q\in L^2(\Omega)$ and $(v_{0m},T_{0m})\rightharpoonup(v_0,T_0)$ in $H,$ let $(v_m(t),T_m(t))$ be a sequence of weak solution for problem \eqref{2.3}-\eqref{2.5} such that $(v_m(0),T_m(0))=(v_{0m},T_{0m}).$ For any $S>0,$ if there exists a subsequence converging ($\ast$-) weakly in spaces $\{(v,T)\in L^{\infty}(0,S;H)\cap L^2(0,S; V):(v_t,T_t)\in L^1(0,S;((H^2(\Omega))^3\cap V)')\}$ to a certain function $(v(t),T(t)).$
Then $(v(t),T(t))$ is a weak solution on $[0,S]$ with $(v(0),T(0))=(v_0,T_0).$
\end{corollary}

\subsection{The existence of a global attractor in $X_\ell$}
In this subsection, we will consider the existence of global attractors for problem \eqref{2.3}-\eqref{2.5} by using the $\ell$-trajectory method. From Theorem \ref{3.1}, we deduce that for any $t>0,$ there exists some $t_0\in (0,t)$ such that $(v(t_0),T(t_0))\in V.$ Therefore, we infer from Remark 2.1 in \cite{jn} that there exists a unique solution of problem \eqref{2.3}-\eqref{2.5} with smoother initial data $(v(t_0),T(t_0)).$ Therefore, many trajectories may start from the same initial data $(v_0,T_0)\in H.$ Denote by $[\chi^{\beta}(\tau,(v_0,T_0))]_{\tau\in[0,\ell]},$ for short $\chi^{\beta}(\tau,(v_0,T_0))(\beta\in\Gamma_{(v_0,T_0)}),$ where $\Gamma_{(v_0,T_0)}$ is the set of indices marking trajectories starting from $(v_0,T_0).$ In the following, we first give the mathematical framework of attractor.
\begin{definition}
Let $\ell$ be a fixed positive constant. Define
\begin{align*}
X_{\ell}=\bigcup_{(v_0,T_0)\in H}\bigcup_{\beta\in\Gamma_{(v_0,T_0)}}\chi^{\beta}(\tau,(v_0,T_0))
\end{align*}
equipped with the topology of $L^2(0,\ell;H).$
\end{definition}
Since $X_\ell\subset\mathcal{C}([0,\ell];H_w),$ it makes sense to talk about the point values of
trajectories. On the other hand, it is not clear whether $X_\ell$ is closed in $L^2(0,\ell;H)$
and hence $X_\ell$ in general is not a complete metric space. In what follows, we first give the definition of some operators.

For any $t\in [0,1],$  we define the mapping $e_t: X_\ell\rightarrow H\times V_I$ by
\begin{align*}
e_t(\chi)=\chi(t\ell)
\end{align*}
for any $\chi\in X_\ell.$

The operators $L_t:X_{\ell}\rightarrow X_{\ell}$ are given by the relation
\begin{align*}
L_t(\chi(\tau,(v_0,T_0)))=(v,T)(t+\tau,(v_0,T_0)),\,\,\,\tau\in[0,\ell]
\end{align*}
for any $\chi(\tau,(v_0,T_0))\in X_\ell,$ where $(v,T)$ is the unique solution of problem \eqref{2.3}-\eqref{2.5} on $[0,\ell+t]$ such that $(v,T)|_{[0,\ell]}=\chi(\tau,(v_0,T_0)),$ we can easily prove the operators $\{L_t\}_{t\geq0}$ is a semigroup on $X_\ell.$

 From the proof of absorbing balls in \cite{jn}, we immediately obtain the following result.
\begin{theorem}\label{3.2.1}
Assume that $Q\in L^2(\Omega).$ Then there exists a positive constant $\rho_1$ satisfying for any bounded subset $B\subset H,$ there exists a time $\tau_1=\tau_1(B)>0$ such that for any weak solutions of problem \eqref{2.3}-\eqref{2.5} with initial data $(v_0,T_0)\in B,$ we have
\begin{align*}
\|v(t)\|^2+\|T(t)\|^2\leq\rho_1
\end{align*}
for any $t\geq \tau_1.$
\end{theorem}
Let
\begin{align*}
B_0=\left\{(v,T)\in V:\|v\|^2+\|T\|^2\leq\rho_1\right\},
\end{align*}
we infer from Theorem \ref{3.2.1} that there exists a time $t_0=t_0(B_0)\geq 0$ such that for any $(v_0,T_0)\in B_0$ and any $t\geq t_0,$ we have
\begin{align*}
(v(t),T(t))\in B_0,
\end{align*}
where $(v(t),T(t))$ is the solution of problem \eqref{2.3}-\eqref{2.5} with initial data $(v_0,T_0)\in B_0.$

Define
\begin{align*}
A(t,(v_0,T_0))=&\left\{(v(t),T(t)):(v(t),T(t))\,\,\textit{is the solution of problem \eqref{2.3}-\eqref{2.5} with initial data} (v_0,T_0)\right\},\\
B_1=&\bigcup\limits_{t\in [0,t_0]}\{A(t,(v_0,T_0)):(v_0,T_0)\in B_0\},\\
B_2=&\overline{B_1}^H
\end{align*}
and
\begin{align*}
B_0^\ell=\{\chi\in X_\ell:e_0(\chi)\in B_2\},
\end{align*}
from the proof of absorbing balls in \cite{jn} and Theorem \ref{3.2.1}, we deduce
\begin{align*}
\{A(t,(v_0,T_0)):(v_0,T_0)\in B_1\}\subset B_1
\end{align*}
for any $t\geq 0$ and $B_1$ is a bounded subset of $V.$ Moreover, we have the following conclusion.
\begin{proposition}\label{3.2.2}
Assume that $B_1$ is a bounded subset of $V.$ Then $B_2=\overline{B_1}^H$ is also a positively invariant, bounded subset of $V.$
\end{proposition}
\textbf{Proof.} From the definition of $B_2,$ we infer that for any $x\in B_2,$ there exists a sequence $\{x_n\}_{n=1}^\infty\subset B_1$ such that
\begin{align*}
x_n\rightarrow x\,\,\textit{in}\,\, H,\,\,\textit{as}\,\,n\rightarrow\infty.
\end{align*}
Since $x_n$ is uniformly bounded in $V$ and $V$ is a reflexive Hilbert space, we deduce that there exist some $y\in V$ and a subsequence $\{x_{n_j}\}_{j=1}^\infty$ of $\{x_n\}_{n=1}^\infty$ such that
\begin{align*}
x_{n_j}\rightharpoonup y\,\,\textit{in}\,\, V,\,\,\textit{as}\,\,j\rightarrow\infty.
\end{align*}
From the compactness of $V\subset H$ and the lower semi-continuity of $\|\cdot\|,$ we obtain
\begin{align*}
x=y
\end{align*}
and
\begin{align*}
\|y\|\leq\lim\inf\limits_{n\rightarrow +\infty}\|x_n\|.
\end{align*}
Therefore, $B_2$ is a bounded subset of $V.$

For any $x\in B_2$ and any fixed $t>0,$ there exists a sequence $\{x_n\}_{n=1}^\infty\subset B_1$ such that $x_n\rightarrow x$ in $H$ as $n\rightarrow\infty,$ we infer from Lemma \ref{3.2} that $A(t,x_n)\rightarrow A(t,x)$ in $H$ as $n\rightarrow\infty.$ Notice that $A(t, x_n)\in B_1$ for any $n\in\mathbb{Z}^+,$ we obtain $A(t,x)\in B_2.$ Therefore, we obtain
\begin{align*}
\{A(t,(v_0,T_0)):(v_0,T_0)\in B_2\}\subset B_2
\end{align*}
for any $t\geq 0.$ \\
\qed\hfill

From Theorem \ref{3.2.1}, we immediately obtain the following result.
\begin{corollary}\label{3.2.3}
Assume that $Q\in L^2(\Omega).$ Then for any bounded subset $B^\ell\subset X_\ell,$ there exists a time $\tau_2=\tau_2(B^\ell)>0$ such that for any weak solutions of problem \eqref{2.3}-\eqref{2.5} with short trajectory $\chi\in B^\ell,$ we have
\begin{align*}
\|v(t)\|^2+\|T(t)\|^2\leq\rho_1
\end{align*}
for any $t\geq \tau_2.$
\end{corollary}

Next, we prove the existence of absorbing sets in $X_\ell$ for the three dimensional viscous primitive equations of large-scale atmosphere.
\begin{theorem}\label{3.2.4}
Assume that $Q\in L^2(\Omega).$ Then there exists a positive constant $\rho_2$ satisfying for the $B_0^\ell,$ there exists a time $\tau_3=\tau_3(B_0^\ell)>0$ such that for any weak solutions of equations \eqref{2.3}-\eqref{2.5} with short trajectory $\chi(\tau,(v_0,T_0))\in B_0^\ell,$ we have
\begin{align*}
\int_0^{\ell}\|(v,T)(t+\tau)\|^2\,d\tau+\left(\int_0^{\ell}\|(v_t,T_t)(t+\tau)\|_{((H^2(\Omega))^3\cap V)'}\,d\tau\right)^2\leq\rho_2
\end{align*}
for any $t\geq \tau_3.$
\end{theorem}
\textbf{Proof.} Taking the $L^2(\Omega)$ inner product of the second equation of \eqref{2.3} with
$T$ and combining Lemma \ref{2.2.1} with Young inequality, we obtain
\begin{align*}
\frac{1}{2}\frac{d}{dt}\|T\|_2^2+\|T\|^2=&\int_\Omega QT\,dxdydz\\
\leq & \|Q\|_2\|T\|_2\\
\leq&\frac{1}{2}\|T\|^2+\frac{1}{2}K_2\|Q\|_2^2,
\end{align*}
which implies that
\begin{align*}
\frac{d}{dt}\|T\|_2^2+\frac{1}{K_2}\|T\|^2_2 \leq K_2\|Q\|_2^2
\end{align*}
and
\begin{align}\label{3.2.5}
\frac{d}{dt}\|T\|_2^2+\|T\|^2\leq K_2\|Q\|_2^2.
\end{align}
It follows from the classical Gronwall inequality that
\begin{align}\label{3.2.6}
\|T(t)\|_2^2 \leq\|T_0\|_2^2e^{-\frac{t}{K_2}}+K_2^2\|Q\|_2^2.
\end{align}
Thanks to
\begin{align}\label{3.2.7}
\frac{d}{ds}(\|T(s)\|_2^2e^{\frac{s}{K_2}})\leq K_2\|Q\|_2^2e^{\frac{s}{K_2}},
\end{align}
for any $\tau\in (0,\ell),$ integrating \eqref{3.2.7} with respect to $s$ from $\tau$ to $t+\tau$ and integrating the resulting inequality over $(0,\ell)$ with respect to $\tau,$ we obtain
\begin{align}\label{3.2.8}
\nonumber\int_0^\ell\|T(t+\tau)\|_2^2\,d\tau\leq&e^{-\frac{t}{K_2}}\int_0^\ell\|T(\tau)\|_2^2\,d\tau+K_2^2\|Q\|_2^2\ell\\
\leq&e^{-\frac{t}{K_2}}(K_2\|T_0\|_2^2+K_2^2\|Q\|_2^2\ell)+K_2^2\|Q\|_2^2\ell.
\end{align}
Multiplying the first equation of \eqref{2.3} by $v$ and integrating over $\Omega,$ we find
\begin{align*}
\frac{1}{2}\frac{d}{dt}\|v\|_2^2+\|v\|^2=&-\int_\Omega\int_0^z\nabla T(x,y,s,t)ds\cdot v(x,y,z,t)\\
\leq &C\|T\|_2\|\nabla v\|_2.
\end{align*}
Let $\lambda=\sup\{\lambda<\frac{1}{K_2}:\lambda\|v\|_2^2\leq\|v\|^2, \forall v\in V_1\},$ we infer from Young inequality and Poinc\'{a}re inequality that
\begin{align}\label{3.2.9}
\frac{d}{dt}\|v\|_2^2+\|v\|^2\leq \frac{C}{\lambda}\|T\|_2^2
\end{align}
and
\begin{align}\label{3.2.10}
\frac{d}{dt}\|v\|_2^2+\lambda\|v\|^2_2\leq \frac{C}{\lambda}\|T\|_2^2.
\end{align}
We infer from the classical Gronwall inequality and \eqref{3.2.6} that
\begin{align}\label{3.2.11}
\nonumber\|v(t)\|_2^2\leq&\|v_0\|_2^2e^{-\lambda t}+\frac{C}{\lambda}\int_0^t\|T(s)\|_2^2e^{\lambda (s-t)}\,ds\\
\nonumber\leq&\|v_0\|_2^2e^{-\lambda t}+\frac{C}{\lambda}\int_0^t\|T_0\|_2^2e^{-\frac{s}{K_2}+\lambda(s-t)}+K_2^2\|Q\|_2^2e^{\lambda (s-t)}\,ds\\
\leq&\|v_0\|_2^2e^{-\lambda t}+\frac{CK_2}{\lambda(1-K_2\lambda)}\|T_0\|_2^2e^{-\lambda t}+\frac{C}{\lambda^2}K_2^2\|Q\|_2^2.
\end{align}
Thanks to
\begin{align}\label{3.2.12}
\frac{d}{ds}(\|v(s)\|_2^2e^{\lambda s})\leq \frac{C}{\lambda}\|T(s)\|_2^2e^{\lambda s}.
\end{align}
Integrating \eqref{3.2.12} with respect to $s$ between $\tau$
and $t+\tau$ and integrating the resulting inequality with respect to $\tau$ over $(0,\ell),$ using
\eqref{3.2.6} and \eqref{3.2.11}, we know
\begin{align}\label{3.2.13}
\int_0^{\ell}\|v(t+\tau)\|_2^2\,d\tau\leq\left(\frac{1}{\lambda}\|v_0\|_2^2+\frac{CK_2}{\lambda^2(1-K_2\lambda)}\|T_0\|_2^2+\frac{K_2^2}{\lambda(1-K_2\lambda)}\|T_0\|_2^2+\frac{C}{\lambda^2}K_2^2\|Q\|_2^2\ell\right)e^{-t\lambda}+\frac{C}{\lambda^2}K_2^2\|Q\|_2^2\ell.
\end{align}
From Proposition \ref{3.2.2}, \eqref{3.2.8} and \eqref{3.2.13}, we deduce that there exists some time $t_0=t_0(B_0^\ell)$ such that
\begin{align}\label{3.2.14}
\int_0^{\ell}\|v(t+\tau)\|_2^2+\|T(t+\tau)\|_2^2\,d\tau\leq\varrho_1
\end{align}
for any $t\geq t_0,$ where
\begin{align*}
\varrho_1=2(1+\frac{C}{\lambda^2})K_2^2\|Q\|_2^2\ell.
\end{align*}
Integrating \eqref{3.2.5} and \eqref{3.2.9} between $t-s$ and $t+\ell$ with $t\geq t_0+\frac{\ell}{2},$ $s\in(0,\frac{\ell}{2})$ and combining \eqref{3.2.14}, we obtain
 \begin{align}\label{3.2.15}
\nonumber\int_0^{\ell}\|v(t+\tau)\|^2+\|T(t+\tau)\|^2\,d\tau
 \leq&2K_2\|Q\|_2^2\ell+\|v(t-s)\|_2^2+\|T(t-s)\|_2^2+\frac{C}{\lambda}\int_{-s}^{\ell}\|T(t+\tau)\|_2^2\,d\tau\\
 \nonumber\leq&2K_2\|Q\|_2^2\ell+\|v(t-s)\|_2^2+\|T(t-s)\|_2^2+\frac{C}{\lambda}\int_0^{\ell+\frac{\ell}{2}}\|T(t-s+\tau)\|_2^2\,d\tau\\
 \leq&2K_2\|Q\|_2^2\ell+\|v(t-s)\|_2^2+\|T(t-s)\|_2^2+\frac{2C}{\lambda}\varrho_1.
 \end{align}
Integrating \eqref{3.2.15} with respect to $s$ over $(0,\frac{\ell}{2})$ and using \eqref{3.2.14}, we find
 \begin{align*}
\frac{\ell}{2}\int_0^{\ell}\|v(t+\tau)\|^2+\|T(t+\tau)\|^2\,d\tau
 \leq&K_2\|Q\|_2^2\ell^2+\int_0^{\frac{\ell}{2}}\|v(t-s)\|_2^2+\|T(t-s)\|_2^2\,ds+\frac{C}{\lambda}\varrho_1\ell\\
 =&K_2\|Q\|_2^2\ell^2+\int_0^{\frac{\ell}{2}}\|v(t-\frac{\ell}{2}+s)\|_2^2+\|T(t-\frac{\ell}{2}+s)\|_2^2\,ds+\frac{C}{\lambda}\varrho_1\ell\\
\leq&K_2\|Q\|_2^2\ell^2+\varrho_1+\frac{C}{\lambda}\varrho_1\ell,
 \end{align*}
 which implies that
 \begin{align}\label{3.2.16}
\int_0^{\ell}\|v(t+\tau)\|^2+\|T(t+\tau)\|^2\,d\tau\leq\varrho_2
 \end{align}
for any $t\geq t_0+\frac{\ell}{2},$ where
\begin{align*}
\varrho_2=2K_2\|Q\|_2^2\ell+(\frac{2C}{\lambda}+\frac{2}{\ell})\varrho_1.
\end{align*}
For any $(\phi,\psi)\in V\cap (H^2(\Omega))^3,$ from H\"{o}lder inequality, we deduce
\begin{align*}
\langle v_t,\phi\rangle\leq &\|v\|_3\|v\|_6\|\nabla\phi\|_2+C\|\nabla v\|_2\|v\|_3\|\phi_z\|_6+C\|v\|_2\|\phi\|_2+\|T\|_2\|\nabla\phi\|_2+\|v\|\|\phi\|,\\
\leq&C(\|v\|^2+1+\|T\|)\|\phi\|_{(H^2(\Omega))^2\cap V_1}
\end{align*}
and
\begin{align*}
\langle T_t,\psi\rangle\leq &\|v\|_3\|T\|_6\|\nabla\psi\|_2+C\|\nabla v\|_2\|T\|_3\|\psi_z\|_6+\|T\|\|\psi\|+\|Q\|_2\|\psi\|_2\\
\leq&C(\|T\|^2+\|v\|^2+\|Q\|_2)\|\psi\|_{H^2(\Omega)\cap V_2},
\end{align*}
which implies that
\begin{align}\label{3.2.17}
\|(v_t,T_t)\|_{(V\cap (H^2(\Omega))^3)'}\leq C(\|v\|^2+\|T\|^2+1+\|Q\|_2).
\end{align}
Integrating \eqref{3.2.17} over $(t,t+\ell)$ and combining \eqref{3.2.16}, we find
\begin{align}\label{3.2.18}
\int_0^{\ell}\|(v_t(t+\tau),T_t(t+\tau))\|_{(V\cap (H^2(\Omega))^3)'}\,d\tau\leq C(\varrho_2+\ell+\ell\|Q\|_2)
\end{align}
for any $t\geq t_0+\frac{\ell}{2}.$\\
\qed\hfill

Let
\begin{align*}
Y=\left\{\chi\in X_\ell:\chi\in L^2(0,\ell; V),\chi_t\in L^1(0,\ell;(V\cap (H^2(\Omega))^3)'\right\}
\end{align*}
equipped with the following norm
\begin{align*}
\|\chi\|_Y=\left\{\int_0^\ell\|\chi(r)\|^2dr+\left(\int_0^\ell\|\chi_t(r)\|_{(V\cap (H^2(\Omega))^3)'}\,dr\right)^2\right\}^{\frac{1}{2}}.
\end{align*}
Define
\begin{align*}
B_1^\ell=\left\{\chi\in X_\ell:\|\chi\|_Y^2\leq\rho_2\right\}.
\end{align*}
From Proposition \ref{3.2.2} and Theorem \ref{3.2.4}, we know that $L_t B_0^\ell\subset B_0^\ell$ for any $t\geq 0$ as well as $L_t B_0^\ell\subset B_1^\ell$ for any $t\geq \tau_3.$
\begin{lemma}\label{3.2.19}
Assume that $Q\in L^2(\Omega).$ Then $\overline{L_t B_0^\ell}^{L^2(0,\ell;H)}\subset B_0^\ell$ for any $t\geq 0.$
\end{lemma}
\textbf{Proof.} Thanks to $L_t B_0^\ell\subset B_0^\ell$ for any $t\geq 0,$ it is enough to prove that
\begin{align*}
\overline{B_0^\ell}^{L^2(0,\ell;H)}\subset B_0^\ell.
\end{align*}
For any $\chi_0\in \overline{B_0^\ell}^{L^2(0,\ell;H)},$ there exists a sequence of trajectories $\chi_n\in B_0^\ell$ such that $\chi_n\rightarrow\chi_0$ in $L^2(0,\ell;H),$ which implies that $e_t(\chi_n)\rightarrow e_t(\chi_0)$ in $H$ for almost all $t\in [0,1].$ Since $e_0(\chi_n)\in B_2$ for any $n\in\mathbb{N},$ there exists a subsequence $\{e_0(\chi_{n_j})\}_{j=1}^{\infty}$ of $\{e_0(\chi_n)\}_{n=1}^{\infty}$ and $(u_0,\phi_0)\in V$ such that $e_0(\chi_{n_j})\rightharpoonup (u_0,\phi_0)$ in $V.$ From the proof of the existence of weak solutions for problem \eqref{2.3}-\eqref{2.5}, we deduce that for any $S>0,$ there exists a subsequence converging ($\ast$-) weakly in spaces $\{(v,T)\in L^{\infty}(0,S;H)\cap L^2(0,S; V):(v_t,T_t)\in L^1(0,S;((H^2(\Omega))^3\cap V)')\}$ to a certain function $(v(t),T(t))$ with $(v(0),T(0))=(v_0,T_0).$ Therefore, we obtain $\chi_0\in X_\ell$ from Corollary \ref{3.3}. It remains to show that $e_0(\chi)\in B_2.$ Since $B_2$ is weakly closed in $V,$ we deduce from the proof of Proposition \ref{3.2.2} that $e_t(\chi_0)\in B_2$ for almost all $t\in [0,1].$ In particular, $e_{t_n}(\chi_0)\in B_2$ for any sequence $t_n$ with $t_n\rightarrow 0.$ From the weak continuity of $\chi_0:[0,\ell]\rightarrow H$ and the weak closedness of $B_2,$ we deduce that $e_0(\chi_0)\in B_2.$ Therefore, we obtain $\chi_0\in B_0^\ell.$\\
\qed\hfill

%
\begin{lemma}\label{3.2.20}
Assume that $Q\in L^2(\Omega).$ Then the mapping $L_t:X_{\ell}\rightarrow X_{\ell}$ is locally Lipschitz continuous on $B_1^\ell$ for all $t\geq0.$
\end{lemma}
\textbf{Proof.} For any fixed $t>0,$ let $(v_1,T_1,p_1)$, $(v_2,T_2,p_2)$
be two solutions for problem \eqref{2.3}-\eqref{2.5} with the initial data
$(v_{1_0},T_{1_0}),$ $(v_{2_0},T_{2_0}),$ respectively. Let $v=v_1-v_2,$ $T=T_1-T_2,$ $p=p_1-p_2,$ from the proof of Theorem 2 in \cite{ccs1}, we conclude
\begin{align}\label{3.2.21}
\frac{d}{dt}(\|v(t)\|_2^2+\|T(t)\|_2^2)\leq\mathbb{L}(t)(\|v(t)\|_2^2+\|T(t)\|_2^2),
\end{align}
where
\begin{align*}
\mathbb{L}(t)=C(\|\nabla v_2\|_2^4+\|\nabla T_2\|_2^4+\|\partial_zv_2\|_2^2\|\nabla\partial_zv_2\|_2^2+\|\partial_zT_2\|_2^2\|\nabla\partial_zT_2\|_2^2).
\end{align*}
Let $s\in(0,\ell)$ and integrating \eqref{3.2.21} from $s$ to $t+s,$ we
obtain
\begin{align}\label{3.2.22}
\|v(t+s)\|_2^2+\|T(t+s)\|_2^2
\leq\int_s^{t+s}\mathbb{L}(r)(\|v(r)\|_2^2+\|T(r)\|_2^2)\,dr+\|v(s)\|_2^2+\|T(s)\|_2^2.
\end{align}
From the classical Gronwall inequality, we deduce
\begin{align}\label{3.2.23}
\nonumber\|v(t+s)\|_2^2+\|T(t+s)\|_2^2\leq&(\|v(s)\|_2^2+\|T(s)\|_2^2)\exp(\int_s^{t+s}\mathbb{L}(r)\,dr)\\
\leq&\mathcal{M}_\ell(t)(\|v(s)\|_2^2+\|T(s)\|_2^2),
\end{align}
where
\begin{align}\label{3.2.24}
\mathcal{M}_\ell(t)=\exp(\int_0^{t+\ell}\mathbb{L}(r)\,dr)
\end{align}
is a finite number depending on $(v_{2_0},T_{1_0})\in B_2$ for any fixed $t>0$ from Remark 2.1 in \cite{jn} and the proof of a priori estimates in \cite{ccs1}.

Integrating \eqref{3.2.23} with respect to $s$ for $0$ to $\ell,$ we obtain
\begin{align}\label{3.2.25}
\int_0^{\ell}\|v(t+s)\|_2^2+\|T(t+s)\|_2^2\,ds\leq \mathcal{M}_\ell(t)\int_0^{\ell}\|v(s)\|_2^2+\|T(s)\|_2^2\,ds,
\end{align}
which implies the mapping $L_t:X_{\ell}\rightarrow X_{\ell}$ is locally Lipschitz continuous on $B_1^\ell$ for all $t\geq0.$\\
\hfill\qed

We can immediately obtain the existence of a global attractor in $X_\ell$ from Lemma \ref{2.2.3} stated as follows.
\begin{theorem}\label{3.2.26}
Assume that $Q\in L^2(\Omega).$ Then the semigroup $\{L_t\}_{t\geq0}$ generated by problem \eqref{2.3}-\eqref{2.5} possesses a global attractor $\mathcal{A}_\ell$ in $X_\ell$ and $e_t(\mathcal{A}_\ell)$ is uniformly bounded in $V$ with respect to $t\in [0,1],$ where
\begin{align*}
e_t(\mathcal{A}_\ell)=\{e_t(\chi):\chi\in\mathcal{A}_\ell\}
\end{align*}
for any $t\in [0,1].$
\end{theorem}

In what follows, we prove the smooth property of the semigroup $\{L_t\}_{t\geq0}$ to estimate the fractal dimension of the global attractor $\mathcal{A}_\ell.$
\begin{theorem}\label{3.2.27}
Assume that $Q\in L^2(\Omega),$ let $\chi^1$ and $\chi^2$ be two short trajectories belonging to $\mathcal{A}_\ell.$ Then there exists a positive constant $\kappa$ independent of $t$ such that for arbitrary $t\geq\ell,$ we have
\begin{align*}
\|L_t\chi^1-L_t\chi^2\|_Y^2\leq\kappa\mathcal{M}_\ell(t)\int_0^{\ell}\|\chi^1(r)-\chi^2(r)\|_2^2\,dr,
\end{align*}
where $\mathcal{M}_\ell(t)$ is given in \eqref{3.2.24}.
\end{theorem}
\textbf{Proof.} For any $\chi^1,$ $\chi^2\in \mathcal{A}_\ell,$ let $(v_1(t+\tau),T_1(t+\tau))=L_t\chi^1,$ $(v_2(t+\tau),T_2(t+\tau))=L_t\chi^2$ and let $v=v_1-v_2,$ $T=T_1-T_2.$ Since $e_t(\chi^1)$ and $e_t(\chi^2)$ is uniformly bounded in $V$ with respect to $t\in [0,1]$ for any $\chi^1,$ $\chi^2\in \mathcal{A}_\ell,$ from the proof of Theorem 2 in \cite{ccs1}, we conclude
\begin{align}\label{3.2.28}
\frac{d}{dt}(\|v(t)\|_2^2+\|T(t)\|_2^2)+\|v(t)\|^2+\|T(t)\|^2\leq\mathbb{L}(t)(\|v(t)\|_2^2+\|T(t)\|_2^2),
\end{align}
where
\begin{align*}
\mathbb{L}(t)=C(\|\nabla v_2\|_2^4+\|\nabla T_2\|_2^4+\|\partial_zv_2\|_2^2\|\nabla\partial_zv_2\|_2^2+\|\partial_zT_2\|_2^2\|\nabla\partial_zT_2\|_2^2).
\end{align*}
For any $t\geq\ell,$ integrating \eqref{3.2.28} from $t-s$ to $t+\ell$ with $s\in[0,\frac{\ell}{2}],$ we conclude
\begin{align*}
&\|v(t+\ell)\|_2^2+\|T(t+\ell)\|_2^2+\int_{t-s}^{t+\ell}\|v(\zeta)\|^2+\|T(\zeta)\|^2\,d\zeta\\
\leq&\int_{t-s}^{t+\ell}\mathbb{L}(\zeta)(\|v(\zeta)\|_2^2+\|T(\zeta)\|_2^2)\,d\zeta+\|v(t-s)\|_2^2+\|T(t-s)\|_2^2.
\end{align*}
It follows from the classical Gronwall inequality that
\begin{align}\label{3.2.29}
\nonumber&\|v(t+\ell)\|_2^2+\|T(t+\ell)\|_2^2+\int_{t-s}^{t+\ell}\|v(\zeta)\|^2+\|T(\zeta)\|^2\,d\zeta\\
\leq&\exp(\int_{t-s}^{t+\ell}\mathbb{L}(\zeta)\,d\zeta)(\|v(t-s)\|_2^2+\|T(t-s)\|_2^2).
\end{align}
For any $t\geq\ell$ and any $s\in[0,\frac{\ell}{2}],$ integrating \eqref{3.2.28} from $s$ to $t-s,$ we obtain
\begin{align*}
\|v(t-s)\|_2^2+\|T(t-s)\|_2^2\leq\int_s^{t-s}\mathbb{L}(r)(\|v(r)\|_2^2+\|T(r)\|_2^2)\,dr+(\|v(s)\|_2^2+\|T(s)\|_2^2).
\end{align*}
We deduce from the classical Gronwall inequality that
\begin{align}\label{3.2.30}
\nonumber\|v(t-s)\|_2^2+\|T(t-s)\|_2^2
\leq&(\|v(s)\|_2^2+\|T(s)\|_2^2)\exp(\int_s^{t-s}\mathbb{L}(r)\,dr)\\
\leq&(\|v(s)\|_2^2+\|T(s)\|_2^2)\exp(\int_0^{t-s}\mathbb{L}(r)\,dr).
\end{align}
Combining \eqref{3.2.29} with \eqref{3.2.30}, we obtain
\begin{align*}
\int_0^\ell\|v(t+\zeta)\|^2+\|T(t+\zeta)\|^2\,d\zeta
\leq&\exp(\int_0^{t+\ell}\mathbb{L}(\zeta)\,d\zeta)(\|v(s)\|_2^2+\|T(s)\|_2^2)\\
=&\mathcal{M}_\ell(t)(\|v(s)\|_2^2+\|T(s)\|_2^2).
\end{align*}
Integrating the above inequality over $(0,\frac{\ell}{2})$ with respect to $s,$ we obtain
\begin{align}\label{3.2.31}
\int_0^\ell\|v(t+\zeta)\|^2+\|T(t+\zeta)\|^2\,d\zeta
\leq\frac{2\mathcal{M}_\ell(t)}{\ell}\int_0^{\frac{\ell}{2}}\|v(s)\|_2^2+\|T(s)\|_2^2\,ds.
\end{align}
Thanks to $\mathcal{M}_\ell(t)$ is bounded for any fixed $t\in [\ell,S],$ we obtain
\begin{align}\label{3.2.32}
\int_0^\ell\|v(t+\zeta)\|^2+\|T(t+\zeta)\|^2\,d\zeta\leq \frac{2\mathcal{M}_\ell(t)}{\ell}\int_0^\ell\|v(s)\|_2^2+\|T(s)\|_2^2\,ds.
\end{align}
Thanks to
\begin{align}\label{3.2.33}
\|v_t\|_{((H^2(\Omega))^2\cap V_1)'}\leq\|v\|+C\|v_1\|_3\|v\|_6+C\|\nabla v_1\|_2\|v\|_3+C\|v_2\|_3\|v\|_6+C\|\nabla v\|_2\|v_2\|_3+C\|T\|_2+C\|v\|_2
\end{align}
and
\begin{align}\label{3.2.34}
\|T_t\|_{(H^2(\Omega)\cap V_2)'}\leq\|T\|+C\|v_1\|_3\|T\|_6+C\|\nabla v_1\|_2\|T\|_3+C\|T_2\|_3\|v\|_6+C\|\nabla v\|_2\|T_2\|_3,
\end{align}
we infer from Theorem \ref{3.2.26}, \eqref{3.2.32}-\eqref{3.2.34} that
\begin{align}\label{3.2.35}
\left(\int_0^\ell\|(v_t(t+r),T_t(t+r))\|_{((H^2(\Omega))^3\cap V)'}\,dr\right)^2\leq \kappa_2\mathcal{M}_\ell(t)\int_0^\ell\|v(s)\|_2^2+\|T(s)\|_2^2\,ds.
\end{align}
The proof of Theorem \ref{3.2.27} is completed.\\
\qed\hfill

From Lemma \ref{2.2.4}, Theorem \ref{3.2.26} and Theorem \ref{3.2.27}, we immediately obtain the following result.
\begin{theorem}\label{3.2.36}
Assume that $Q\in L^2(\Omega).$ Then the fractal dimension of the global attractor $\mathcal{A}_\ell$ in $X_\ell$ of the semigroup $\{L_t\}_{t\geq0}$ generated by problem \eqref{2.3}-\eqref{2.5} established in Theorem \ref{3.2.26} is finite.
\end{theorem}
\subsection{The existence of a global attractor in $H$}
In this subsection, we prove the existence of a finite dimensional global attractor in $H$ of problem \eqref{2.3}-\eqref{2.5}. From Lemma \ref{3.2}, we deduce that for any given initial condition $(v_0,T_0)\in B_2\subset V,$ there exists a unique solution of problem \eqref{2.3}-\eqref{2.5}, hence solution operators $S(t)$ is a semigroup on $B_2.$ Moreover, $B_2$ is positively invariant with respect to $S(t).$
\begin{theorem}\label{4.2.1}
Assume that $Q\in L^2(\Omega).$ Then the mapping $e_1:\mathcal{A}_\ell\rightarrow \mathcal{A}=e_1(\mathcal{A}_\ell)$ is Lipschitz continuous. That is, for any two short trajectories $\chi^1,$ $\chi^2\in\mathcal{A}_\ell,$ there exists a positive constant $\theta$ dependent on $\ell$ such that
\begin{align*}
\|e_1(\chi^1)-e_1(\chi^2)\|_2^2\leq\theta\int_0^{\ell}\|\chi^1(r)-\chi^2(r)\|_2^2\,dr.
\end{align*}
\end{theorem}
\textbf{Proof.} For any $\chi^1,$ $\chi^2\in \mathcal{A}_\ell,$ let $(v_1(t+\tau),T_1(t+\tau))=L_t\chi^1,$ $(v_2(t+\tau),T_2(t+\tau))=L_t\chi^2$ and let $v=v_1-v_2,$ $T=T_1-T_2.$ Thanks to $e_0(\chi^1)$ and $e_0(\chi^2)$ is uniformly bounded in $V$ for any $\chi^1,$ $\chi^2\in \mathcal{A}_\ell,$ from \eqref{3.2.28}, we obtain
\begin{align*}
\frac{d}{dt}(\|v(t)\|_2^2+\|T(t)\|_2^2)+\|v(t)\|^2+\|T(t)\|^2\leq\mathbb{L}(t)(\|v(t)\|_2^2+\|T(t)\|_2^2).
\end{align*}
For $s\in (0,\ell),$ we infer from the classical Gronwall inequality that
\begin{align}\label{4.2.2}
\nonumber\|v(\ell)\|_2^2+\|T(\ell)\|_2^2\leq&(\|v(s)\|_2^2+\|T(s)\|_2^2)\exp(\int_s^\ell\mathbb{L}(r)\,dr)\\
\leq&(\|v(s)\|_2^2+\|T(s)\|_2^2)\exp(\int_0^\ell\mathbb{L}(r)\,dr).
\end{align}
Integrating \eqref{4.2.2} over $(0,\ell),$ we obtain
\begin{align*}
\|v(\ell)\|_2^2+\|T(\ell)\|_2^2
\leq\frac{1}{\ell}\exp(\int_0^\ell\mathbb{L}(r)\,dr)\int_0^\ell(\|v(s)\|_2^2+\|T(s)\|_2^2)\,ds.
\end{align*}
Thanks to \eqref{3.2.24}, we know that
\begin{align*}
\mathcal{M}_\ell=\exp(\int_0^\ell\mathbb{L}(r)\,dr)<+\infty,
\end{align*}
which implies that the mapping $e_1:\mathcal{A}_\ell\rightarrow \mathcal{A}$ is Lipschitz continuous.\\
\hfill\qed
\begin{theorem}\label{4.2.3}
Assume that $Q\in L^2(\Omega).$ Then dynamical system $(S(t),B_2)$ possesses a global attractor $\mathcal{A}=e_1(\mathcal{A}_\ell)$ in $H,$ which is
compact, invariant in $H$ and attracts every bounded subset in $H$ with respect
to the topology of $H.$ Furthermore, $\mathcal{A}$ is bounded in $V$ and its fractal dimension is finite.
\end{theorem}
\textbf{Proof.} From Lemma \ref{2.2.5}, Theorem \ref{3.2.36} and Theorem \ref{4.2.1}, we know that $\mathcal{A}$ is compact and the fractal dimension of $\mathcal{A}$ is finite. As a result of $L_t\mathcal{A}_\ell=\mathcal{A}_\ell,$ we have
\begin{align*}
S(t)\mathcal{A}=S(t)e_1(\mathcal{A}_\ell)=e_1(L_t\mathcal{A}_\ell)=e_1(\mathcal{A}_\ell)=\mathcal{A}
\end{align*}
for any $t\geq 0.$ From the definition of $B_2,$ we deduce that for any bounded subset of $H,$ there exists some time $\bar{t}=\bar{t}(B)$ such that for any $t\geq \bar{t},$ we have
\begin{align*}
\{A(t,(v_0,T_0)):(v_0,T_0)\in B\}\subset B_2.
\end{align*}
Therefore, we only need to prove that
\begin{align*}
\lim_{t\rightarrow+\infty}dist_H(S(t)B_2,\mathcal{A})=0.
\end{align*}
Otherwise, there exist some positive constant $\epsilon_0,$ some sequence $\{(v_n,T_n)\}_{n=1}^{\infty}\subset B_2$ and some $\{t_n\}_{n=1}^{\infty}$ with $t_n\rightarrow+\infty$ as $n\rightarrow+\infty$ such that
\begin{align}\label{4.2.4}
dist_H(S(t_n)(v_n,T_n),\mathcal{A})\geq\epsilon_0.
\end{align}
From the definition of $B_2,$ we deduce that there exists $\chi_n\in B_0^\ell$ such that
\begin{align*}
(v_n,T_n)=e_0(\chi_n).
\end{align*}
Since $\{\chi_n\}_{n=1}^{\infty}$ is bounded in $X_\ell$ and $\mathcal{A}_\ell$ is a global attractor in $X_\ell$ of the semigroup $\{L_t\}_{t\geq 0}$ generated by problem \eqref{2.3}-\eqref{2.5}, there exist a subsequence $\{\chi_{n_j}\}_{n=1}^{\infty}$ of $\{\chi_n\}_{n=1}^{\infty}$ and a subsequence $\{t_{n_j}\}_{n=1}^{\infty}$ of $\{t_n\}_{n=1}^{\infty}$ such that
\begin{align*}
L_{t_{n_j}-\ell}\chi_{n_j}\rightarrow\chi\in\mathcal{A}_\ell\,\,\,\textit{in}\,\,\, X_\ell\,\,\,\textit{for}\,\,\textit{as}\,\,\,j\rightarrow+\infty.
\end{align*}
Thanks to the continuity of $e_1,$ we have
\begin{align*}
S(t_{n_j})(v_{n_j},T_{n_j})=e_1(L_{t_{n_j}-\ell}\chi_{n_j})\rightarrow e_1(\chi)\in\mathcal{A}\,\,\,\textit{in}\,\,\, H\,\,\,\textit{as}\,\,\,j\rightarrow+\infty,
\end{align*}
which contradicts with \eqref{4.2.4}.\\
\qed\hfill

Since $\mathcal{A}$ is bounded in $V,$ we immediately obtain the following result.
\begin{theorem}\label{4.2.5}
Assume that $Q\in L^2(\Omega)$ and $\mathcal{A}_1$ is the global attractor established in \cite{jn}. Then
\begin{align*}
\mathcal{A}_1=\mathcal{A}.
\end{align*}
\end{theorem}

\section*{Acknowledgement}
 This work was supported by the National Science Foundation of China Grant (11401459).

\bibliographystyle{elsarticle-template-num}
\bibliography{BIB}

\begin{thebibliography}{34}
\expandafter\ifx\csname natexlab\endcsname\relax\def\natexlab#1{#1}\fi
\providecommand{\bibinfo}[2]{#2}
\ifx\xfnm\relax \def\xfnm[#1]{\unskip,\space#1}\fi
\bibitem{ccs3}
\bibinfo{author}{C.~S. Cao}, \bibinfo{author}{J.~K. Li}, \bibinfo{author}{E.~S.
  Titi},
\newblock \bibinfo{title}{Global well-posedness of strong solutions to the 3d
  primitive equations with horizontal eddy diffusivity},
\newblock \bibinfo{journal}{Journal of Differential Equations}
  \bibinfo{volume}{257(11)} (\bibinfo{year}{2014}{\natexlab{a}})
  \bibinfo{pages}{4108--4132}.
\bibitem{ccs4}
\bibinfo{author}{C.~S. Cao}, \bibinfo{author}{J.~K. Li}, \bibinfo{author}{E.~S.
  Titi},
\newblock \bibinfo{title}{Local and global well-posedness of strong solutions
  to the 3d primitive equations with vertical eddy diffusivity},
\newblock \bibinfo{journal}{Archive for Rational Mechanics and Analysis}
  \bibinfo{volume}{214(1)} (\bibinfo{year}{2014}{\natexlab{b}})
  \bibinfo{pages}{35--76}.
\bibitem{ccs1}
\bibinfo{author}{C.~S. Cao}, \bibinfo{author}{E.~S. Titi},
\newblock \bibinfo{title}{Global well-posedness of the three-dimensional
  viscous primitive equations of large-scale ocean and atmosphere dynamics},
\newblock \bibinfo{journal}{Annals of Mathematics} \bibinfo{volume}{166}
  (\bibinfo{year}{2007}) \bibinfo{pages}{245--267}.
\bibitem{ccs2}
\bibinfo{author}{C.~S. Cao}, \bibinfo{author}{E.~S. Titi},
\newblock \bibinfo{title}{Global well-posedness of the 3d primitive equations
  with partial vertical turbulence mixing heat diffusion},
\newblock \bibinfo{journal}{Communications in Mathematical Physics}
  \bibinfo{volume}{310} (\bibinfo{year}{2012}) \bibinfo{pages}{537--568}.
\bibitem{cvv}
\bibinfo{author}{V.~Chepyzhov}, \bibinfo{author}{M.~Vishik},
  \bibinfo{title}{Attractors for equations of mathematical physics},
  \bibinfo{publisher}{American Mathematical Society, Providence, RI},
  \bibinfo{year}{2002}.
\bibitem{ci}
\bibinfo{author}{I.~Chueshov},
\newblock \bibinfo{title}{A squeezing property and its applications to a
  description of long-time behaviour in the three-dimensional viscous primitive
  equations},
\newblock \bibinfo{journal}{Proceedings of the Royal Society of Edinburgh:
  Section A Mathematics} \bibinfo{volume}{144(4)} (\bibinfo{year}{2014})
  \bibinfo{pages}{711--729}.
\bibitem{elc}
\bibinfo{author}{L.~C. Evans}, \bibinfo{author}{R.~Gastler},
\newblock \bibinfo{title}{Some results for the primitive equations with
  physical boundary conditions},
\newblock \bibinfo{journal}{Zeitschrift f\:{u}r angewandte Mathematik und
  Physik} \bibinfo{volume}{64(6)} (\bibinfo{year}{2013})
  \bibinfo{pages}{1729--1744}.
\bibitem{ebd}
\bibinfo{author}{B.~D. Ewaldy}, \bibinfo{author}{R.~Temam},
\newblock \bibinfo{title}{Maximum principles for the primitive equations of the
  atmosphere},
\newblock \bibinfo{journal}{Discrete and Continuous Dynamical Systems A}
  \bibinfo{volume}{7} (\bibinfo{year}{2001}) \bibinfo{pages}{343--362}.
\bibitem{gf}
\bibinfo{author}{F.~Guill\'{e}n-Gonz\'{a}lez}, \bibinfo{author}{N.~Masmoudi},
  \bibinfo{author}{M.~A. Rodr\'{i}guez-Bellido},
\newblock \bibinfo{title}{Anisotropic estimates and strong solutions for the
  primitive equations},
\newblock \bibinfo{journal}{Differential and Integral Equations}
  \bibinfo{volume}{14} (\bibinfo{year}{2001}) \bibinfo{pages}{1381--1408}.
\bibitem{gbl2}
\bibinfo{author}{B.~L. Guo}, \bibinfo{author}{D.~W. Huang},
\newblock \bibinfo{title}{Existence of weak solutions and trajectory attractors
  for the moist atmospheric equations in geophysics},
\newblock \bibinfo{journal}{Journal of mathematical physics}
  \bibinfo{volume}{47} (\bibinfo{year}{2006}) \bibinfo{pages}{1089--7658}.
\bibitem{gbl}
\bibinfo{author}{B.~L. Guo}, \bibinfo{author}{D.~W. Huang},
\newblock \bibinfo{title}{On the existence of atmospheric attractors},
\newblock \bibinfo{journal}{Science in China D} \bibinfo{volume}{51}
  (\bibinfo{year}{2008}) \bibinfo{pages}{469--480}.
\bibitem{gbl1}
\bibinfo{author}{B.~L. Guo}, \bibinfo{author}{D.~W. Huang},
\newblock \bibinfo{title}{On the 3d viscous primitive equations of the
  large-scale atmosphere},
\newblock \bibinfo{journal}{Acta Mathematica Scientia B} \bibinfo{volume}{29}
  (\bibinfo{year}{2009}) \bibinfo{pages}{846--866}.
\bibitem{gbl3}
\bibinfo{author}{B.~L. Guo}, \bibinfo{author}{D.~W. Huang},
\newblock \bibinfo{title}{Existence of the universal attractor for the $3-d$
  viscous primitive equations of large-scale moist atmosphere},
\newblock \bibinfo{journal}{Journal of Differential Equations}
  \bibinfo{volume}{251} (\bibinfo{year}{2011}) \bibinfo{pages}{457--491}.
\bibitem{hng}
\bibinfo{author}{N.~G. Holtz}, \bibinfo{author}{I.~Kukavica},
  \bibinfo{author}{V.~Vicol}, \bibinfo{author}{M.~Ziane},
\newblock \bibinfo{title}{Existence and regularity of invariant measures for
  the three dimensional stochastic primitive equations},
\newblock \bibinfo{journal}{Journal of Mathematical Physics}
  \bibinfo{volume}{55} (\bibinfo{year}{2014}) \bibinfo{pages}{pp:051504}.
\bibitem{hch}
\bibinfo{author}{C.~H. Hsiaa}, \bibinfo{author}{M.~C. Shiueb},
\newblock \bibinfo{title}{Time-periodic solutions of the primitive equations of
  large-scale moist atmosphere: existence and stability},
\newblock \bibinfo{journal}{Applicable Analysis} \bibinfo{volume}{94(9)}
  (\bibinfo{year}{2015}) \bibinfo{pages}{1926--1963}.
\bibitem{hc1}
\bibinfo{author}{C.~Hu},
\newblock \bibinfo{title}{Asymptotic analysis of the primitive equations under
  the small depth assumption},
\newblock \bibinfo{journal}{Nonlinear Analysis} \bibinfo{volume}{61}
  (\bibinfo{year}{2005}) \bibinfo{pages}{425--460}.
\bibitem{hc}
\bibinfo{author}{C.~Hu}, \bibinfo{author}{R.~Temam},
  \bibinfo{author}{M.~Ziane},
\newblock \bibinfo{title}{The primitive equations of the large scale ocean
  under the small depth hypothesis},
\newblock \bibinfo{journal}{Discrete and Continuous Dynamical Systems. A}
  \bibinfo{volume}{9} (\bibinfo{year}{2003}) \bibinfo{pages}{97--131}.
\bibitem{jn}
\bibinfo{author}{N.~Ju},
\newblock \bibinfo{title}{The global attractor for the solutions to the three
  dimensional viscous primitive equations},
\newblock \bibinfo{journal}{Discrete and Continuous Dynamical Systems A}
  \bibinfo{volume}{17} (\bibinfo{year}{2007}) \bibinfo{pages}{159--179}.
\bibitem{jn2}
\bibinfo{author}{N.~Ju},
\newblock \bibinfo{title}{The finite dimensional global attractor for the 3d
  viscous primitive equations},
\newblock \bibinfo{journal}{Discrete and Continuous Dynamical Systems}
  \bibinfo{volume}{36(12)} (\bibinfo{year}{2016}) \bibinfo{pages}{7001--7020}.
\bibitem{jn1}
\bibinfo{author}{N.~Ju}, \bibinfo{author}{R.~Temam},
\newblock \bibinfo{title}{Finite dimensions of the global attractor for 3d
  primitive equations with viscosity},
\newblock \bibinfo{journal}{Journal of Nonlinear Science}
  \bibinfo{volume}{25(1)} (\bibinfo{year}{2015}) \bibinfo{pages}{131--155}.
\bibitem{lk}
\bibinfo{author}{K.~Li}, \bibinfo{author}{F.~Li},
\newblock \bibinfo{title}{Pullback attractor for nonautonomous primitive
  equations of large-scale ocean and atmosphere dynamics},
\newblock \bibinfo{journal}{Abstract and Applied Analysis}
  \bibinfo{volume}{2013} (\bibinfo{year}{2013}) \bibinfo{pages}{Article ID
  691615, 12 pages}.
\bibitem{ljl}
\bibinfo{author}{J.~L. Lions}, \bibinfo{author}{R.~Temam},
  \bibinfo{author}{S.~Wang},
\newblock \bibinfo{title}{New formulations of the primitive equations of
  atmosphere and applications},
\newblock \bibinfo{journal}{Nonlinearity} \bibinfo{volume}{5}
  (\bibinfo{year}{1992}{\natexlab{a}}) \bibinfo{pages}{237--288}.
\bibitem{ljl1}
\bibinfo{author}{J.~L. Lions}, \bibinfo{author}{R.~Temam},
  \bibinfo{author}{S.~Wang},
\newblock \bibinfo{title}{On the equations of the large-scale ocean},
\newblock \bibinfo{journal}{Nonlinearity} \bibinfo{volume}{5}
  (\bibinfo{year}{1992}{\natexlab{b}}) \bibinfo{pages}{1007--1053}.
\bibitem{mj}
\bibinfo{author}{J.~M\'{a}lek}, \bibinfo{author}{J.~Ne\u{c}as},
\newblock \bibinfo{title}{A finite-dimensional attractor for three-dimensional
  flow of incompressible fluids},
\newblock \bibinfo{journal}{Journal of Differential Equations}
  \bibinfo{volume}{127} (\bibinfo{year}{1996}) \bibinfo{pages}{498--518}.
\bibitem{mj1}
\bibinfo{author}{J.~M\'{a}lek}, \bibinfo{author}{D.~Pra\u{z}\'{a}k},
\newblock \bibinfo{title}{Large time behavior via the method of
  $\ell$-trajectories},
\newblock \bibinfo{journal}{Journal of Differential Equations}
  \bibinfo{volume}{181} (\bibinfo{year}{2002}) \bibinfo{pages}{243--279}.
\bibitem{mtt}
\bibinfo{author}{T.~T. Medjo},
\newblock \bibinfo{title}{On the uniqueness of $z$-weak solutions of the
  three-dimensional primitive equations of the ocean},
\newblock \bibinfo{journal}{Nonlinear Analysis} \bibinfo{volume}{11}
  (\bibinfo{year}{2010}) \bibinfo{pages}{1413 --1421}.
\bibitem{pm}
\bibinfo{author}{M.~Petcu}, \bibinfo{author}{R.~Temam},
  \bibinfo{author}{D.~Wirosoetisno},
\newblock \bibinfo{title}{Existence and regularity results for the primitive
  equations in two space dimensions},
\newblock \bibinfo{journal}{Communications on Pure and Applied Analysis}
  \bibinfo{volume}{3} (\bibinfo{year}{2004}) \bibinfo{pages}{115--131}.
\bibitem{rjc}
\bibinfo{author}{J.~C. Robinson}, \bibinfo{title}{Infinite-dimensional
  Dynamical Systems}, \bibinfo{publisher}{Cambridge University Press},
  \bibinfo{year}{2001}.
\bibitem{sj2}
\bibinfo{author}{J.~Simon},
\newblock \bibinfo{title}{Compact sets in the space $l^p(0,t;b)$},
\newblock \bibinfo{journal}{Annali di Matematica Pura ed Applicata}
  \bibinfo{volume}{146} (\bibinfo{year}{1987}) \bibinfo{pages}{65--96}.
\bibitem{tr}
\bibinfo{author}{R.~Temam}, \bibinfo{title}{Infinite-Dimensional Dynamical
  Systems in Mechanics and Physics}, \bibinfo{publisher}{New York,
  Springer-Verlag}, \bibinfo{year}{1997}.
\bibitem{tr1}
\bibinfo{author}{R.~Temam}, \bibinfo{author}{M.~Ziane},
\newblock \bibinfo{title}{Some mathematical problems in geophysical fluid
  dynamics},
\newblock \bibinfo{journal}{Handbook of Mathematical Fluid Dynamics}
  \bibinfo{volume}{3} (\bibinfo{year}{2005}) \bibinfo{pages}{535--658}.
\bibitem{yb2}
\bibinfo{author}{B.~You}, \bibinfo{author}{S.~Ma},
\newblock \bibinfo{title}{Global attractors for three dimensional viscous
  primitive equations of large-scale atmosphere in log-pressure coordinate},
\newblock \bibinfo{journal}{Abstract and Applied Analysis}
  \bibinfo{volume}{2013} (\bibinfo{year}{2013}) \bibinfo{pages}{Article ID
  758730, 16 pages}.
\bibitem{zmc1}
\bibinfo{author}{M.~C. Zelati}, \bibinfo{author}{M.~Fremond},
  \bibinfo{author}{R.~Temam}, \bibinfo{author}{J.~Tribbia},
\newblock \bibinfo{title}{The equations of the atmosphere with humidity and
  saturation:uniqueness and physical bounds},
\newblock \bibinfo{journal}{Physica D} \bibinfo{volume}{264}
  (\bibinfo{year}{2013}) \bibinfo{pages}{49--65}.
\bibitem{zmc}
\bibinfo{author}{M.~C. Zelati}, \bibinfo{author}{A.~M. Huang},
  \bibinfo{author}{I.~Kukavica}, \bibinfo{author}{R.~Temam},
  \bibinfo{author}{M.~Ziane},
\newblock \bibinfo{title}{The primitive equations of the atmosphere in presence
  of vapor saturation},
\newblock \bibinfo{journal}{Nonlinearity} \bibinfo{volume}{28}
  (\bibinfo{year}{2015}) \bibinfo{pages}{625--668}.

\end{thebibliography}
\end{document}